\documentclass[aps,prl,10pt,twocolumn,superscriptaddress]{revtex4-1}

\usepackage[verbose=true]{microtype}
\usepackage{amsmath}
\usepackage{amssymb}
\usepackage{quantum}
\usepackage{graphicx}
\usepackage{subfig}
\usepackage{color}
\usepackage[hypertexnames=false,pdfpagelabels]{hyperref}
\usepackage[amsmath,thmmarks,hyperref]{ntheorem}
\usepackage[capitalise,poorman]{cleveref}

\theoremstyle{break}

\theoremstyle{break}
\theorembodyfont{\normalfont}

\theoremstyle{nonumberplain}
\theorembodyfont{\normalfont}
\theoremsymbol{$\Box$}

\DeclareMathAlphabet{\mathitbf}{OML}{cmm}{b}{it}

\newcommand{\chan}{\mathcal{E}}
\newcommand{\cN}{\mathcal{N}}

\newcommand{\keyword}[1]{\emph{#1}}
\newcommand{\flip}{\mathbb{F}}

\begin{document}

\title{Entanglement can completely defeat quantum noise}

\author{Jianxin Chen}
\affiliation{Department of Mathematics \& Statistics, U.\ of Guelph,
  Guelph, Ontario, Canada}
\affiliation{Institute for Quantum Computing, U.\ of Waterloo,
  Waterloo,  Ontario, Canada}
\author{Toby S. Cubitt}
\affiliation{Departamento de An\'alisis Matem\'atico, U.\ Complutense
  de Madrid, Plaza de Ciencias 3, 28040 Madrid, Spain}
\author{Aram W. Harrow}
\affiliation{
 Department of Computer Science \& Engineering, U.\ of Washington,
 Seattle, WA 98195, USA}
\author{Graeme Smith}
\affiliation{%
  IBM T.J.\ Watson Research Center, Yorktown Heights, NY 10598, USA}

\begin{abstract}
  We describe two quantum channels that individually cannot send any
  information, even classical, without some chance of decoding error. But
  together a single use of each channel can send quantum information perfectly
  reliably. This proves that the zero-error classical capacity exhibits
  \emph{superactivation}, the extreme form of the superadditivity
  phenomenon in which entangled inputs allow communication over zero capacity
  channels. But our result is stronger still, as it even allows zero-error
  \emph{quantum} communication when the two channels are combined. Thus our
  result shows a new remarkable way in which entanglement across two systems
  can be used to resist noise, in this case perfectly. We also show a new form
  of superactivation by entanglement shared between sender and receiver.
\end{abstract}


\maketitle

Sending information over a noisy communication channel usually requires error
correction. 
The best transmission rate possible, optimized over all conceivable
error-correction strategies, is called the capacity of the channel. The
capacity tells us the value of a noisy channel for communication and is
measured in bits per channel use. Capacities are central to the theory of
information initiated by Shannon \cite{Shannon48}, and serve as guideposts for
the development of practical communication schemes.

The usual setting for information theory is the asymptotic regime, where
sender and receiver have many independent uses of a fixed noisy channel. The
probability of transmission error is required to vanish in the limit of many
channel uses, and the resulting capacity is called the Shannon capacity. A
more demanding requirement is to insist that the error probability be
\emph{exactly} zero. This leads to zero-error information theory, also studied
by Shannon~\cite{Shannon_zero-error}. The zero-error setting has a more
combinatorial flavor; indeed, a large part of modern graph theory owes its
origins to the study of zero-error communication~\cite{zero-error_review}.
Zero-error information theory is most relevant when the asymptotic guarantees
of Shannon theory are insufficient---either because the number of channel uses
isn't large enough to make the probability of error small, or because
absolutely no error can be tolerated. Furthermore, it is related to the rate
at which the error probability tends to zero in the usual Shannon
capacity~\cite{SGB67}.

Ultimately, noisy communication links are described by quantum mechanics, and
in systems such as optical communication, quantum effects cannot be neglected.
When considering the zero-error capacity of quantum channels, we may consider
either the classical or quantum capacities, measuring respectively the rate at
which a channel may send bits or qubits without any error. The resulting
coding problems lead to rich generalizations of the graph theory problems
arising from classical channels~\cite{DSW10,*Beigi10}.

Even classically, the zero-error capacity is quite different from the Shannon
capacity. For example, it is non-additive~\cite{Shannon_zero-error}. However,
some basic properties are common to both capacities. One of the most basic is
the behavior of zero-capacity channels, i.e.\ channels that are too noisy to
transmit any information. It seems like combining two such completely useless
channels will still not allow communication. Indeed, for classical channels,
this intuition is correct. The only classical channels with no zero-error
capacity are those where every pair of inputs have some non-zero probability
of being confused at the output (otherwise we could use a non-confusable pair
of inputs to send one bit). But if we use two such channels together, any pair
of inputs to the combined channel can also be confused, so the joint channel
has zero capacity.

Remarkably, we show that this elementary property of classical channels fails
for quantum channels: there are pairs of quantum channels, each with no
classical zero-error capacity at all, yet which \emph{do} allow perfect
classical transmission when the two channels are combined. This striking
phenomenon is known as \keyword{superactivation}~\cite{graeme+jon,*SSY11}, and
to our knowledge this is the first superactivation of a classical capacity of
standard quantum channels. In fact, we can strengthen this result to show that
the joint channel can even transmit far more delicate \emph{quantum}
information with zero error, so these channels superactivate both the
classical and quantum zero-error capacity \emph{simultaneously}, an extreme
form of superactivation never before seen.

It is a little like having two pipes, both completely blocked, allowing
nothing to flow through them. Yet, by plumbing the blocked pipes in parallel,
water can flow. Of course, this analogy breaks down due to quantum effects.
Indeed, entanglement is at the heart of this remarkable superactivation
phenomenon. Our results show that using input states entangled across the two
inputs to the joint channel, we can completely defeat noise, allowing perfect
transmission where none would otherwise be possible. Finally, we also show
that entanglement can be used to completely defeat noise in a new form of
superactivation: superactivation by entanglement, where entanglement between
the sender and receiver allows perfect communication with a zero-capacity
channel.

\paragraph{Related phenomena.}
Superactivation has previously been found only for the quantum capacity of
quantum channels~\cite{graeme+jon,*SSY11}. By contrast, the classical capacity
is nonzero for any nontrivial quantum channel, so superactivation is trivially
impossible in the standard Shannon setting. There, the possibility of
superadditivity (two channels having greater asymptotic capacity when
combined) remains a major open question, since the additivity violation of
Hasting~\cite{Hastings} only addressed one-shot (Holevo) capacities. Private
communication is intermediate between classical and quantum communication.
Here, superadditivity has been observed~\cite{private1,*private2}, while
superactivation remains an open question. In the zero-error case, other
researchers have found activation results for single copies of quantum
channels~\cite{Runyao}. It has also been shown that the zero-error capacity
even of classical channels can be increased by shared entanglement between
sender and receiver, though it cannot be superactivated~\cite{CLMW09,*CLMW10}.
Recently, others have started to explore how quantum zero-error capacities
relate to the graph-theoretic quantities that provide bounds on zero-error
capacities of classical channels~\cite{DSW10,*Beigi10}.

\paragraph{Overview of technical contributions.}
We use two key technical ideas. The first is to choose our channels as
randomly as possible, subject to certain constraints. The constraints
guarantee that combining the two channels allows information to be
transmitted. Meanwhile, the random choice helps ensure that the individual
channels are noisy, so transmit very little information. This
\emph{constrained-randomness} strategy has had many applications in quantum
information. Examples include \cite{HLW06}, which considered random states
subject to a rank constraint; \cite{Andreas+Patrick,Hastings}, which chose
random channels subject to a constraint ensuring their product gave one large
output eigenvalue; or the variant in \cite{rank_additivity}, which choose
random channels subject to a constraint that guaranteeing a singular output
for an appropriate entangled input.

This strategy is very successful in showing that there is a nonzero
probability of picking a channel that is noisy enough on a single copy. This
is sufficient for all the above results, as they only concern ``one-shot''
quantities: properties of a single copy of a channel. But we need something
stronger; we require \emph{arbitrarily many} copies of the \emph{same} channel
to be so noisy that the channel has no zero-error capacity (even
asymptotically). We only get to exploit a finite amount of randomness (in the
choice of one copy of the channel), yet we want this finite amount of
randomness to control a property of an arbitrarily large and highly correlated
object (the capacity of arbitrarily many copies of that same channel). However
large the probability of picking a suitable channel for a single copy, unless
that probability is 1 it will shrink 
to zero on a growing number of copies. So, on its own, this strategy
completely fails to give results for asymptotic quantities such as channel
capacities.

Our second technical tool is a new method of controlling the behavior of an
unbounded number of copies of the channel through randomness on a single copy
using algebraic geometry. Such arguments show that certain bad sets (say, the
set of channels for which $k$ copies can send a classical bit with zero error)
have zero measure, so that even a union of countably many of them does as
well. In other words, we show that the probability of picking a suitable
constrained random channel is exactly~1, thus avoiding any decay in the
probability for growing numbers of copies. These techniques rely on a greater
knowledge of the structure of the problem, but are still highly general, and
we suspect they will have further application in quantum information,
including problems in which small errors are tolerated.

\paragraph{Proof of main result.}
We now describe the proof of zero-error superactivation. Recall that two
quantum states $\rho,\sigma$ are perfectly distinguishable exactly when they
are orthogonal ($\tr[\rho\sigma]=0$). Thus, the classical zero-error capacity
of a channel $\chan$ is 0 exactly when no pair of inputs gives orthogonal
outputs. Mathematically, we require
\begin{equation}\label{eq:individual}
  \forall\psi,\varphi \quad \tr[\chan(\varphi)\chan(\psi)] \neq 0.
\end{equation}
Let $\circ$ denote composition, define $\chan^*$ by $\tr A\chan(B) =
\tr\chan^*(A)B$ and $\cN:=\chan^*\circ\chan$. Then \cref{eq:individual} is
equivalent to requiring $\forall \psi,\varphi\,\tr[\varphi\cN(\psi)]\neq 0$.
This in turn is equivalent to insisting that the (CP, but not necessarily
trace preserving) map $\cN$ always has full rank output. This condition was
previously used in \cite{rank_additivity} to find multiplicativity violations
for the minimum output rank. Following \cite{rank_additivity}, we can rewrite
this condition in terms of the Choi matrix\footnote{The Choi matrix of a map
  $\cN$ is
  $\sigma_{AB} = (\cN\otimes\id)(\proj{\omega})$, where $\ket{\omega} =
  \sum_i\ket{i}\ket{i}/\sqrt{d}$ and $\id$ is the identity map.} $\sigma_{AB}$
of the composite map $\cN=\chan^*\circ\chan$ (recalling that the action of the
map can be recovered from the Choi matrix via $\cN(\rho) =
\tr[\sigma_{AB}\cdot(\rho^T\ox\id)]$), to obtain
$\tr[\sigma_{AB}\cdot(\psi\ox\varphi)] \neq 0$. In other words, the support of
$\sigma_{AB}$ (denoted $S_{AB}$) contains no product
states.\footnote{Ref~\cite{rank_additivity} refers to channels, but the
  arguments extend to $\cN=\chan^*\ox\chan$, even if it is not a channel. But
  the arguments go through unchanged for arbitrary linear maps.} The same
argument holds for any number of copies $k$ of the channel. So, for a channel
to have no zero-error capacity even asymptotically, $S_{AB}^{\ox k}$ must
contain no product states \emph{for any tensor power} $k$ (unlike
\cite{rank_additivity}, where $k=1$ sufficed).

Furthermore, in contrast to~\cite{rank_additivity}, it is no longer true that
\emph{any} bipartite subspace $S_{AB}$ will suffice; the fact that the
subspace must now support the Choi matrix of a composite map $\cN =
\chan^*\circ\chan$ imposes extra symmetry requirements on $S_{AB}$. It is easy
to verify that $S_{AB}$ must have the following additional properties:
(i)~$\flip(S_{AB}) = S_{AB}$, where $\flip(\sum_{ij}\alpha_{ij}\ket{i}\ket{j})
= \alpha_{ij}^*\ket{j}\ket{i}$ swaps the two systems and takes complex
conjugates, (ii)~$S_{AB}$ contains a state $\ket[AB]{\psi} =
\sum_{i,j}\alpha_{ij}\ket[A]{i}\ket[B]{j}$ whose matrix of coefficients
$M=[\alpha_{ij}]$ is positive-definite (has strictly positive eigenvalues).
These two conditions are also sufficient, in the following sense: given a
subspace $S_{AB}$ satisfying (i) and (ii), one can always construct a Choi
matrix $\sigma_{AB}$ supported on $S_{AB}$ which corresponds to some map $\cN
= \chan^*\circ\chan$. To see this, choose a (not necessarily orthonormal)
basis $\ket{\psi_k}$ for $S_{AB}$ whose coefficient matrices $M_k$ are
positive-definite (condition (ii) guarantees that such a basis exists).
Denoting the eigenvectors of $M_k$ by $\ket{\phi_i^k}$, the matrix
\begin{equation}
  \rho_{AB}
  = \sum_{ijk}\ket[A]{\phi_i^k}\ket[B]{k,i}\bra[A]{\phi_j^k}\bra[B]{k,j}.
\end{equation}
is (up to rescaling) a Choi matrix for a channel $\chan$, such that the Choi
matrix of $\cN = \chan^*\circ\chan$ is supported on $S_{AB}$.

To prove superactivation, we need a \emph{pair} of channels $\chan_1$ and
$\chan_2$, each satisfying \cref{eq:individual} so that it has no zero-error
capacity, but such that the joint channel $\chan_1\ox\chan_2$ \emph{does} have
positive capacity. For the latter, we need a pair of input states that are
mapped to orthogonal outputs by the joint channel, i.e.
\begin{equation}\label{eq:joint}
  \exists\psi,\varphi \quad
  \tr[(\chan_1\ox\chan_2)(\psi)\cdot(\chan_1\ox\chan_2)(\varphi)] = 0,
\end{equation}
as we can use these states to perfectly transmit 1~bit.
Generalising~\cite{rank_additivity}, we choose these inputs $\psi,\varphi$ to
be maximally entangled states $\ket{\omega} = \sum_i\ket{i}\ket{i}/\sqrt{d}$
and $\ket{\omega'}=(X\ox\id)\ket{\omega}$, where $X$ is the unitary consisting
of 1's down the anti-main diagonal (i.e.\ the generalisation to arbitrary
dimension of the Pauli matrix $\sigma_x$). Rewriting \cref{eq:joint} in terms
of the Choi matrices $\sigma_{1,2}$ of the composite maps
$\cN_{1,2}=\chan_{1,2}^*\circ\chan_{1,2}$, we obtain for this choice of input
states that the Choi matrices must satisfy
$\tr[\sigma_1^T\cdot(X\ox\id)\sigma_2(X\ox\id)] = 0$. But, denoting the
supports of $\sigma_{1,2}$ by $S_{1,2}$, this simply states that the subspaces
$S_1$ and $(X\ox\id) S_2$ should be orthogonal. We might as well take $S_2 =
(X\ox\id)S_1^\perp$, since this still allows zero-error communication with the
composite channel, while making $S_2$ as large as possible can only help
suppress the single-use capacity.

Our task now reduces to finding an appropriate $S_1$, which we call simply $S$
from now on. To summarize the constraints described above, we require:
\begin{enumerate}
\item[(i)] $(S^{\otimes k})^\perp$ contains no product states for any $k$;
\item[(ii)] $((S^\perp)^{\otimes k})^\perp$ contains no product states for any
  $k$;
\item[(iii)] $\flip(S) = S$;
\item[(iv)] $\flip((X\ox\id)\cdot S) = (X\ox\id)\cdot S$;
\item[(v)] $S$ contains a state with positive-definition coefficient matrix;
\item[(vi)] $(X\ox\id)\cdot S$ contains a state with positive-definition
  coefficient matrix.
\end{enumerate}
Properties (iii)\nobreakdash--(vi) guarantee that $S_1=S$ and
$S_2=(X\ox\id)S^\perp$ correspond to valid channels. Our choice of $S_2$
ensures that these channels together can communicate one bit without error.
Most of the remaining work is showing that a random $S$ satisfies
(i)\nobreakdash--(ii): arbitrary tensor powers contain no product states,
ensuring that the individual channels have no zero-error capacity. (In stating
property (ii), we have used the fact that the set of product states is left
invariant by $(X\ox\id)$.) A priori, this appears extremely demanding, since
we must satisfy an infinite number of constraints simultaneously; indeed, we
only get to choose a subspace from a constant number of dimensions, but we
need to rule out product states on an unbounded number of tensor copies.
However, algebraic geometry arguments will show, remarkably, almost \emph{all}
subspaces (all but a measure-zero set) satisfy properties
(i)\nobreakdash--(ii).

\begin{figure}[!htbp]
  \centering
  \subfloat[\label{fig:everything}]{%
    \includegraphics[scale=0.6]{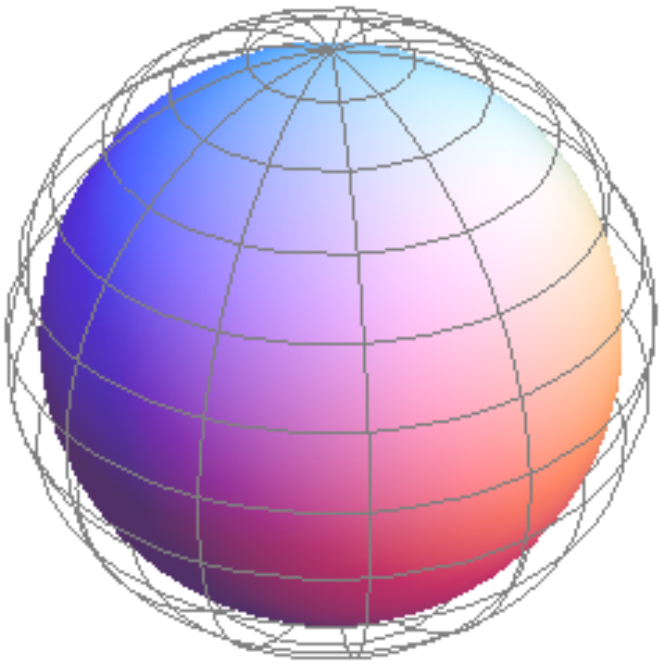}}
  \quad
  \subfloat[\label{fig:nothing}]{%
    \includegraphics[scale=0.6]{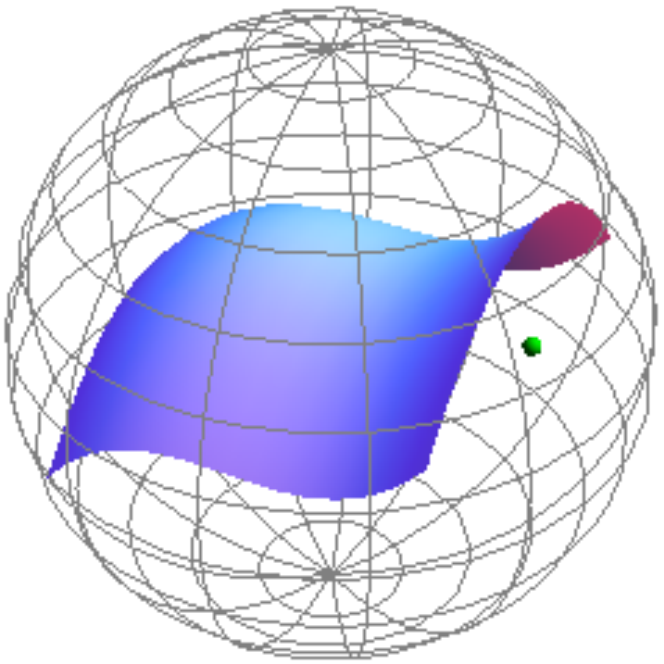}}
  \caption{The set of zeros of a set of simultaneous polynomial equations
    either \subref{fig:everything}~comprises the entire space, or
    \subref{fig:nothing}~has measure~0. To show that it is measure~0, it
    therefore suffices to find a single point outside of the
    set~\subref{fig:nothing}.}
  \label{fig:Zariski_dichotomy}
\end{figure}

There is a standard way to represent a subspace as a
vector \footnote{This is known as the \keyword{Pl\"ucker embedding}.
}, writing $S$ as the antisymmetric product of an orthonormal basis of $S$
(e.g.\ consider the unique state of $\dim S$ fermions with state space $S$ ).
With this parameterisation, we can see that the set $E_k$ of all subspaces
whose $k^{\text{th}}$ tensor power \emph{does} contain a product state is
given by a set of simultaneous homogenous polynomial equations~\cite{CCH09}.
These are the subspaces that we \emph{don't} want; we are looking for a
subspace that is \emph{not} in any $E_k$. In general, the set of zeros of a
set of polynomials will either have measure~0, or will comprise the entire
space (\cref{fig:Zariski_dichotomy}) \footnote{This is because if the zero set
  of a polynomial contains a set with non-empty interior, 
  then by Taylor expansion that polynomial must be zero everywhere.}.

Therefore, to show that $E_k$ has zero measure, it suffices to show that there
is a \emph{single} point outside of it, thereby ruling out the possibility
that it is the entire space. To do this we construct such a subspace from an
unextendible product basis (UPB): a basis on a bipartite space which cannot be
extended by adding any further orthogonal product states. Now, the orthogonal
complement of the span of a UPB is by definition a subspace that contains no
product states and, since the tensor product of two UPBs is again a
UPB~\cite{CCH09}, this is also true for any tensor power of this subspace, as
required. This subspace will certainly not satisfy the other requirements
(ii)\nobreakdash--(vi) (in particular, its orthogonal complement is just the
span of the UPB, which clearly \emph{does} contain product states,
dramatically failing to satisfy (ii)). But its existence shows that there is a
subspace that is not contained in any $E_k$, which is sufficient to collapse
each $E_k$ to a set of zero measure (\cref{fig:Zariski_dichotomy}). Thus
$\cup_{k\geq 1} E_k$ has zero measure, so property (i) holds for a random $S$
with probability 1. Since $S^\perp$ is also uniformly random, property (ii)
automatically holds with probability 1 as well.

Our argument must be refined to handle (iii--vi). Start with (iii) and (iv),
which are linear constraints on the subspace, and thus translate into
polynomial constraints in the coordinates parameterising the subspace. Since
the intersection of the solutions to two sets of simultaneous polynomial
equations is the set defined by the union of both sets of polynomials, we can
use the preceding argument \emph{within} this intersection. The only
modification is that we now choose a UPB that satisfies the symmetry
requirements. This is achieved by symmetrizing an arbitrary UPB. Since adding
additional product states to a UPB maintains the UPB property, this requires
only that the initial UPB is not too large, which holds already for the UPBs
from~\cite{Bhat04}. Therefore, we can choose a random subspace satisfying
requirements (iii) and (iv) and with probability~1 it will also satisfy
requirements (i) and (ii).

Finally, we address (v) and (vi). These are not algebraic, so we cannot repeat
the algebraic geometry argument. However, the requirement that our bipartite
subspace contain a state with positive-definite coefficient matrix is quite
mild. In particular, if a subspace has this property and we perturb it by a
sufficiently small amount, then the positive-definite element stays positive
definite. So, around every such subspace there is an open ball of subspaces
that also satisfy the positivity requirement. Therefore, the set of subspaces
satisfying (v) and (vi) has positive measure, even relative to the set of
subspaces satisfying (iii) and (iv). We have seen that the set of subspaces
satisfying (i) and (ii) is full measure within the set of subspaces satisfying
(iii) and (iv). The intersection of a positive-measure set with a full-measure
set has positive measure, so the set of subspaces satisfying (i) to (vi) has
positive measure. So, at least one such subspace $S$ must exist. We have
already shown that this is equivalent to the existence of a channel that
superactivates the classical zero-error capacity, so we are done.

Armed with these techniques, we can extend our result to show the joint
channel can even transmit \emph{quantum} information with zero
error~\cite{SuperDuper}. Thus, we want the joint channel to transmit at least
one qubit perfectly, meaning that some two-dimensional subspace is transmitted
undisturbed. For this, it is sufficient~\cite{SuperDuper} to find \emph{two}
different pairs of orthogonal input states in the same two-dimensional
subspace are mapped to orthogonal output states. This just adds another
algebraic symmetry condition on $S$, so we can deal with it exactly as before.

\paragraph{Application to entanglement-assisted capacity.}
We have shown that by encoding the information into states $\ket{\omega}$ and
$(X\ox\id)\ket{\omega}$ which are entangled across the two inputs to the joint
channel, we can completely defeat the noise in the channel, even though we
couldn't send \emph{any} information through either channel on its own.
Entanglement can be used to completely defeat noise in another way, by sharing
the entanglement between sender and receiver. To see this, note that we have a
channel $\chan_1$ above with no zero-error capacity, even asymptotically, but
for which inputs $\ket{\psi} = \ket{\omega}$, $\ket{\varphi} =
(X\ox\id)\ket{\omega}$ to the joint channel $\chan_1\ox\chan_2$ give
orthogonal outputs:
$\tr[\chan_1\ox\chan_2(\psi)\cdot\chan_1\ox\chan_2(\varphi)]=0$
(\cref{eq:joint}). But applying a channel cannot increase the orthogonality of
two states, so $\chan_1\ox I(\psi)$ and $\chan_1\ox I(\varphi)$ are
orthogonal. Therefore, if the sender and receiver share the maximally
entangled state $\ket{\omega}$, then the sender can communicate one bit
perfectly to the receiver, by either sending her half of the entangled state
directly through the channel, or first applying the local unitary $X$ to her
half of the state before sending it. Since the resulting states $\chan_1\ox
I(\psi)$ and $\chan_1\ox I(\varphi)$ are orthogonal, they can be perfectly
distinguished by the receiver, thereby transmitting one bit with zero error.

\paragraph{Acknowledgments.}
JC is supported by NSERC. TSC was supported by a Leverhulme early career
fellowship, the EU (QAP, QESSENCE, MINOS, COMPAS, COQUIT, QUEVADIS), and
Spanish grants QUITEMAD, I-MATH, and MTM2008-01366. AWH was funded by NSF
grants 0916400, 0829937, 0803478, DARPA QuEST contract FA9550-09-1-0044 and
the IARPA MUSIQC and QCS contracts. GS is supported by the DARPA QuEST
contract HR0011-09-C-0047.

\bibliography{zero-error_PRL}

\end{document}